\documentclass[twocolumn,showpacs,preprintnumbers,amsmath,amssymb,superscriptaddress]{revtex4}

\usepackage{epsf}
\usepackage{graphicx}
\usepackage{sidecap}

\usepackage{color}

\def \beq {\begin{equation}}
\def \eeq {\end{equation}}
\begin{document}
 
\title{{Dirac  state switching in transition metal diarsenides}}
\author{Gyanendra~Dhakal}\affiliation {Department of Physics, University of Central Florida, Orlando, Florida 32816, USA}
\author{M.~Mofazzel~Hosen}\affiliation {Department of Physics, University of Central Florida, Orlando, Florida 32816, USA}
 \author{Wei-Chi Chu} \affiliation {Department of Physics, Northeastern University, Boston, Massachusetts 02115, USA}
 \author{Bahadur Singh} \affiliation{Department of Physics, Northeastern University, Boston, Massachusetts 02115, USA}
\author{Klauss~Dimitri}\affiliation {Department of Physics, University of Central Florida, Orlando, Florida 32816, USA}
\author{BaoKai Wang}\affiliation {Department of Physics, Northeastern University, Boston, Massachusetts 02115, USA} 
\author{Firoza~Kabir}\affiliation {Department of Physics, University of Central Florida, Orlando, Florida 32816, USA}
\author{Christopher Sims}\affiliation {Department of Physics, University of Central Florida, Orlando, Florida 32816, USA}.
\author{Sabin Regmi} \affiliation {Department of Physics, University of Central Florida, Orlando, Florida 32816, USA}
\author{William Neff} \affiliation {Department of Physics, University of Central Florida, Orlando, Florida 32816, USA}
\author{Dariusz Kaczorowski} \affiliation{Institute of Low Temperature and Structure Research, Polish Academy of Sciences, 50-950 Wroclaw, Poland}
\author{Arun Bansil} \affiliation {Department of Physics, Northeastern University, Boston, Massachusetts 02115, USA}
\author{Madhab~Neupane}
\affiliation {Department of Physics, University of Central Florida, Orlando, Florida 32816, USA}
 
\date{\today}

\begin{abstract}{
 {Topological Dirac and Weyl semimetals, which support low-energy quasiparticles in condensed matter physics, are currently attracting intense interest due to  exotic physical properties such as large magnetoresistance and high carrier mobilities. Transition metal diarsenides such as MoAs$_{2}$ and WAs$_{2}$ have been reported to harbor very high magnetoresistance suggesting the possible existence of a topological quantum state, although this conclusion remains dubious. Here, based on systematic angle-resolved photoemission spectroscopy (ARPES) measurements and parallel first-principles calculations, we investigate the electronic properties of TAs$_{2}$ (T = Mo, W). Importantly,  clear evidence for  switching  the single-Dirac cone surface state in MoAs$_{2}$ with the cleaving plane is observed, whereas a Dirac state is not observed in WAs$_{2}$ despite its high magnetoresistance.  Our study thus reveals the key role of the terminated plane in a low-symmetry system, and provides a new perspective on how termination can drive dramatic changes in electronic structures.}}
\end{abstract}

\maketitle
\noindent
\textbf{INTRODUCTION}\\
The experimental discovery of a 3D topological insulator state is widely acknowledged as a major milestone in condensed matter and materials physics due to its unique properties and potential applications \cite{Hasan, Moore, SCZhang}. This discovery has spurred an intense research interest in other classes of topological materials \cite{ Neupane3, bansil}.  As a consequence, exotic states such as the Dirac semimetal \cite{ Murakami, Young, Wang, Neupane, Nagaosa, MH2,  Young_Kane}, Weyl semimetal \cite{wtrs,TaAs_theory,TaAs_theory_1, Suyang_Science, Hong_Ding, Ilya_PRL, Hasan_2}, nodal-line semimetal \cite{ PbTaSe2, MH3, Schoop, MH1} have been theoretically and experimentally realized. In Dirac semimetals, linearly dispersing bands cross each other at low-binding energies  behaving like Dirac fermions in condensed matter systems. Since inversion and time-reversal symmetries protect the Dirac semimetallic state, its realization requires the material to be centro-symmetric  and non-magnetic \cite{Ashvin, Ilya_review}. Breaking  either inversion or the time-reversal symmetry drives a Dirac semimetal into a Weyl semimetal, so that either  a non-centrosymmetric or a magnetic material is required  to realize the Weyl semimetallic state \cite{Ashvin, Ilya_review}.
Weyl fermions in condensed matter evolve as low-energy excitations at topologically protected crossing points (Weyl nodes) between electron and hole bands. Weyl nodes act like monopoles or antimonopoles of Berry curvature depending on their chirality, and nodes of opposite chiralities are separated in momentum space and behave like a magnetic dipole \cite{wtrs, Ashvin, Ilya_review}. A recent new player on the scene is a new type of Weyl semimetal, called type-II Weyl semimetal, which does not conserve Lorentz invariance in that its Weyl cones are tilted and the electron- and hole-like pockets intersect at the Fermi surface \cite{typeII}. Open Fermi surfaces, anisotropic chiralities, and unconventional anomalous Hall effect are among the unusual features of type-II Weyl fermions. Most type-II Weyl fermions are associated with gigantic magnetoresistance \cite{Ashvin, Ilya_review}. Angle-resolved photoemission spectroscopy  studies with parallel first-principles calculations on   MoTe$_{2}$, WTe$_{2}$, MoP$_{2}$, and WP$_{2}$ \cite{sun, huang, Wu, PRL117}  confirm their type-II Weyl characteristics. These advances have spurred the exploration of type-II Weyl semimetallic states in other binary compounds of Mo and W. Transition metal diarsenides have drawn special attention due to their high magnetoresistance and carrier density-of-states \cite{Jwang}.\\
\begin{figure*}[hbt!]
  \centering
  \includegraphics[width=16cm]{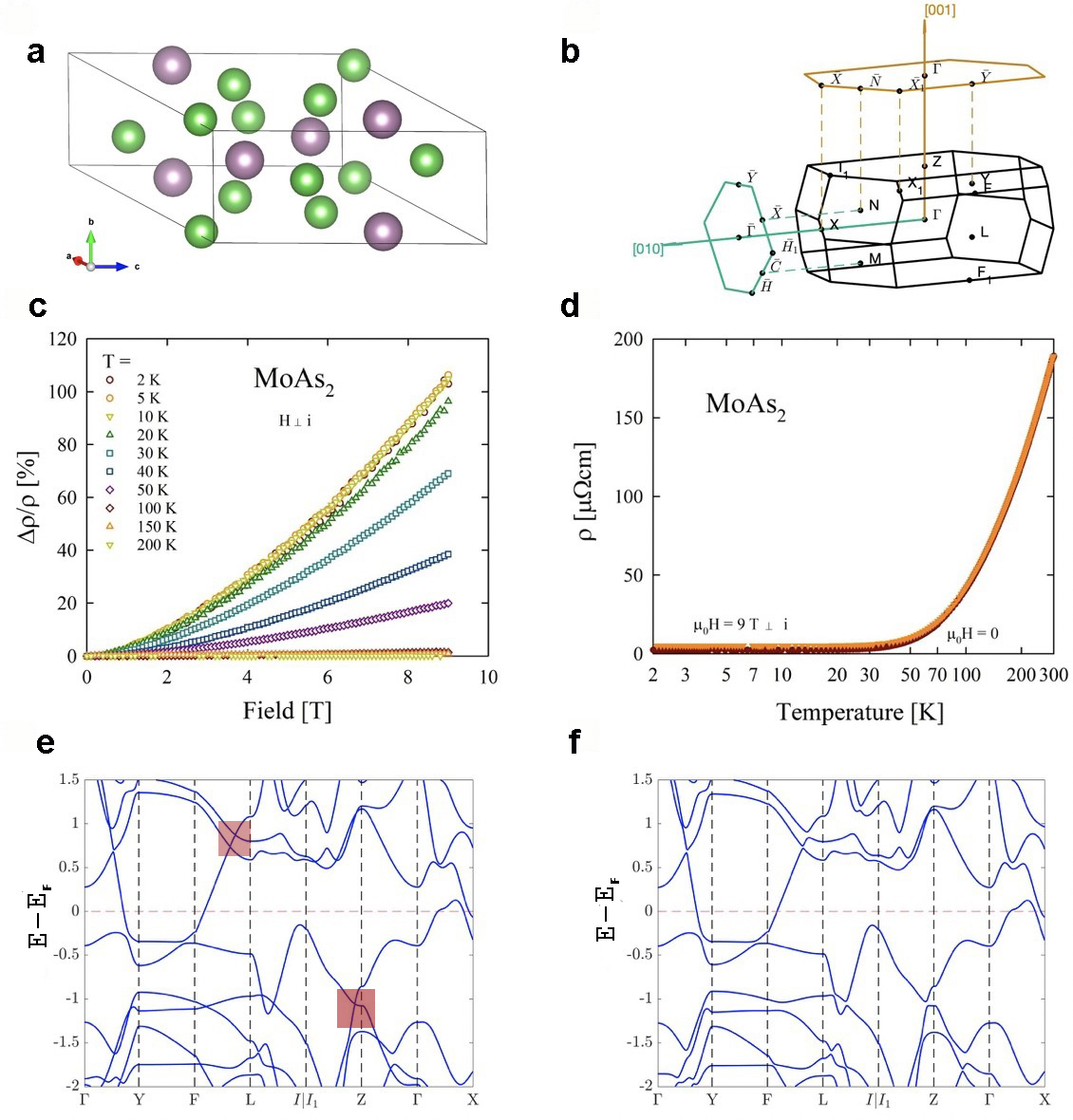}
	\caption{Crystal structure, transport measurements, and bulk band structure of MoAs$_{2}$.
\textbf{a}. Conventional unit cell of MoAs$_{2}$. Purple and green spheres refer to Mo and As atoms, respectively. \textbf{b}. Conventional 3D Brillouin zone of MoAs$_{2}$ \indent with high-symmetry points labeled. Orange and green lines represent the conventional [001] and [010] surface projection, respectively. \textbf{c}. Magnetoresistance as a function of  the perpendicular field at different temperatures. \textbf{d}. Resistivity as a function of  temperature in  MoAs$_{2}$ samples without field (dark red colored symbols) and with 9 T field (orange colored symbols) perpendicular to the current. \textbf{e.} Bulk band structure of  MoAs$_{2}$ without  SOC. Linearly crossed bands are enclosed in the red rectangles.  \textbf{f}. Bulk band structure of  MoAs$_{2}$ with SOC.}

\end{figure*}
\begin{figure*}[hbt!]
	\centering
	\includegraphics[width=18cm]{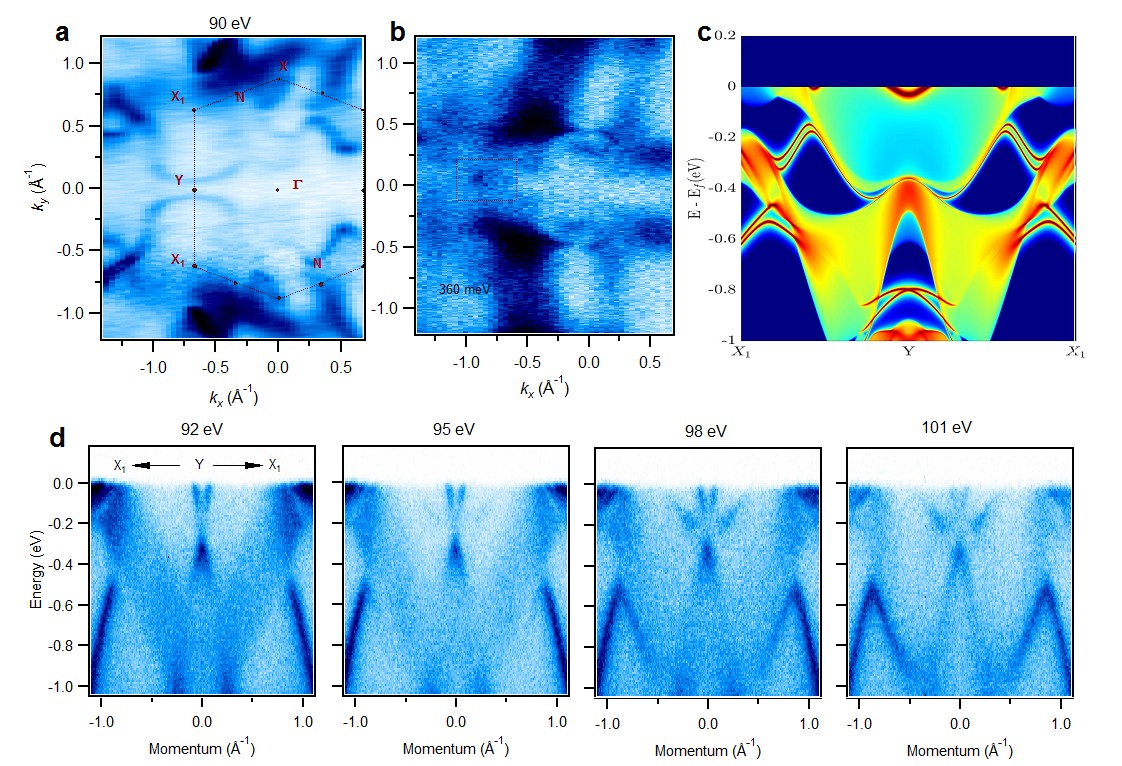}
	\caption{   Electronic structure of MoAs$_{2}$ in the [001] plane.
\textbf{a}. Experimental  Fermi surface map  with high-symmetry points of the Brillouin zone are marked. The dashed lines represent the constructed Brillouin zone. \textbf{b}. Energy contour at the binding energy of 360 meV. The dotted rectangle encloses the Dirac node.
 \textbf{c}. Computed dispersion map along the X$_{1}$-Y-X$_{1}$ direction.  \textbf{d}. Photon energy dependent dispersion measurements along the X$_{1}$-Y-X$_{1}$ direction.  All data were collected at the ALS beamline 4.0.3 at a temperature of 20 K.}
\end{figure*}
Recent reports on the topological properties and the origin of high magnetoresistance in MoAs$_{2}$ present contradictory results \cite{Chen, Lou, Singha}. A first-principles calculation finds states with non-trivial $\mathbb{Z}_{2}$ index and suggests high magnetoresistance to result from electron-hole compensation \cite{Chen}. However, another experimental/ theoretical study of MoAs$_{2}$ finds trivial band structures and argues that the high magnetoresistance results from carrier motion on the Fermi surface, which is enriched via the open-orbit topology \cite{Lou}. A quantum oscillation experiment complemented by first principles calculations indicate the existence  of  open-orbit in MoAs$_{2}$ \cite{Singha}. The need for a comprehensive experimental and theoretical analysis of the electronic structure of MoAs$_{2}$ is thus clear. 
A new level of understanding of the electronic structure can also be expected through the delineation of the termination dependence of the electronic structure that has not yet been reported. To date, ARPES measurements on the related compound WAs$_{2}$ have not been reported, which belongs to the same crystal group as MoAs$_{2}$ but with a stronger spin-orbit coupling (SOC) strength.

Motivated by the preceding considerations, here we report a systematic angle-resolved photoemission spectroscopy (ARPES) study of TAs$_{2}$ (T = Mo, W). We show that the Dirac cone states on the [001] plane in MoAs$_{2}$ switch to become trivial bands on the [010] plane, thus highlighting the role of the cleavage plane in the topology of the states. Interestingly, we find high magnetoresistance and  similar Fermi surface features on the [001] surface in WAs$_{2}$, however WAs$_{2}$ does not harbor Dirac states on the [001] surface. Since MoAs$_{2}$ and WAs$_{2}$ both exhibit metallic band structures at low temperatures with highly populated trivial metallic surface states, our study indicates that topological semimetallic states are not the key for generating high magnetoresistance in these materials. \\
 \begin{figure*}[hbt!]
	\centering
	\includegraphics[width=16cm]{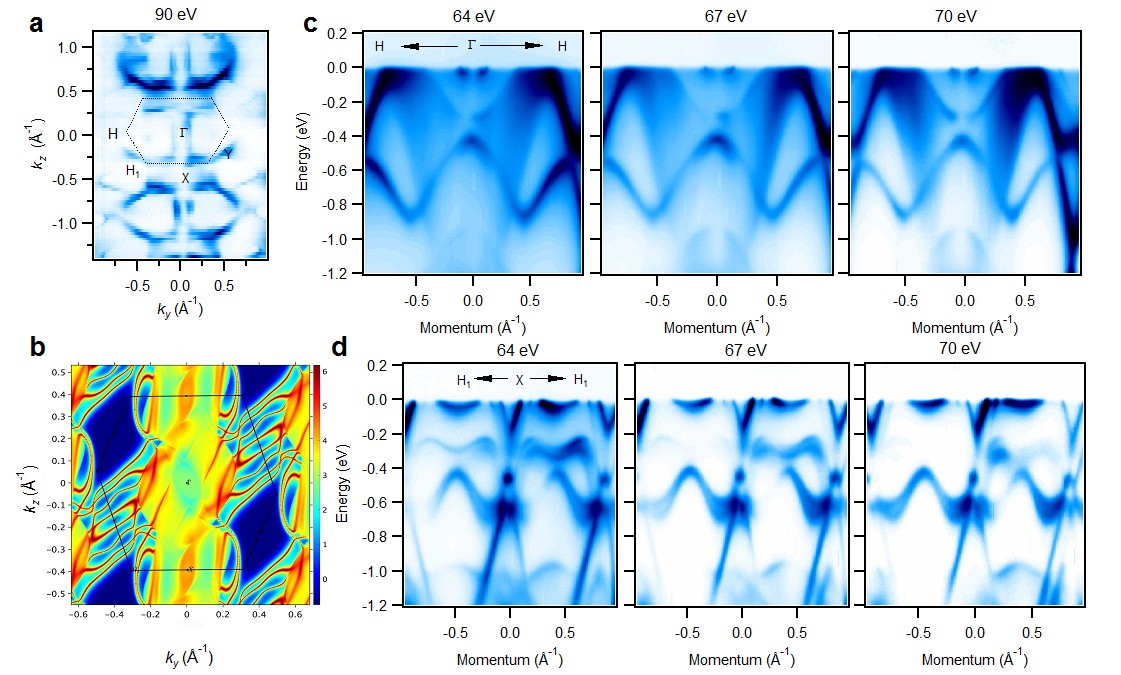}
	\caption{Electronic structures of MoAs$_{2}$ on the [010] surface.
\textbf{a}. Measured Fermi surface map of the MoAs$_{2}$ on the [010] plane at a photon energy of 90 eV. The dashed line represent the constructed Brillouin zone. \textbf{b}. Calculated Fermi surface on the  [010] plane.  \textbf{c}. Photon energy dependent electronic band dispersion along the H-$\Gamma$-H direction, as labeled in the constructed Brillouin zone. \textbf{d}. Photon energy dependent electronic band dispersion along the H$_{1}$-X-H$_{1}$ direction. The data were collected at the ALS beamline 10.0.1  at a temperature of 15 K.}
\end{figure*}

 \noindent
  \textbf{RESULTS }\\
  \noindent
  \textbf{Crystal structure and sample characterization} \\
We begin our discussion with the crystal structure of MoAs$_{2}$, it crystallizes to form a monoclinic crystal structure in the C$_{12/m1}$ symmetry group (\#12),  with  lattice parameters   a = 9.064(7) \AA, b = 3.987(1) \AA, c =7.7182(9) \AA, and  $\beta$= 119.37(1)$^{\circ}$  \cite{J, Jwang}, as shown in Fig. 1a.  The conventional bulk Brillouin zone and its 2D projection onto the [010] and [001] surfaces can be seen in Fig. 1b depicting a hexagonal shape on both projections.  Magnetoresistance   as a function of the field is presented in Fig. 1c for different temperatures noted in the graph, the moderately high magnetoresistance  at low temperature implies the possibility of MoAs$_{2}$ hosting topological states and/or harboring electron-hole compensation. The transport behavior of the sample at low temperature can be associated with  a multi-band system which possesses variable carrier densities as it deviates from the Kohler's rule \cite{Jwang}. The resistivity of MoAs$_{2}$ as a function of the temperature is presented in Fig. 1d covering  a large temperature range from 2 K to 300 K plotted on a semi-logarithmic scale. The resistivity at 300 K reaches nearly 190 $\mu\Omega$cm. The graph clearly shows the metallic nature at low temperature at zero field with a plateau ranging to about 50 K, and a semimetallic behavior above 50 K. The  applied magnetic field hardly alters the resistivity of the sample. The bulk band calculation of  MoAs$_{2}$ without and  with the inclusion of the spin-orbit coupling  is presented in Figs. 1e-f, respectively.   The bands in the vicinity of the Fermi level give an impression of the Dirac dispersion at some points enclosed in  the red rectangles (0.8 eV above the Fermi level near L, 1 eV below the Fermi level near  Z point). Furthermore, the Dirac-like crossings take place away from the high-symmetry points. In addition, they are not robust against spin-orbit coupling as gaps are opened upon the inclusion of  spin-orbit coupling. Therefore, they do not encompass the Dirac semimetallic state featured by bulk Dirac bands in this compound.

\begin{figure*}[hbt!]
	\centering
	\includegraphics[width=16cm]{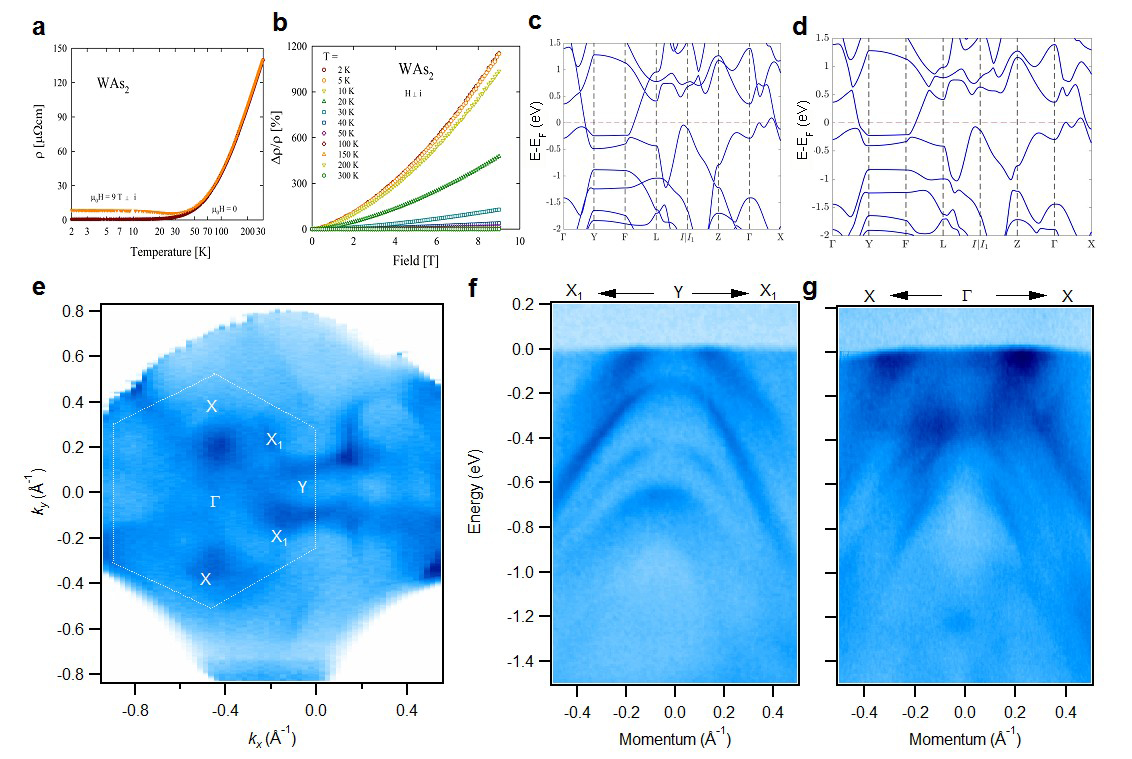}
	\caption{Transport properties and the electronic structure of WAs$_{2}$. \textbf{a}. Resisitivity as a function of  temperature at zero field (dark red colored curve) and under 9 T (orange colored curve). \textbf{b}. Magnetoresistance as a function of the field at different temperatures. \textbf{c}.  Bulk band structure of  WAs$_{2}$ without SOC. \textbf{d}. Bulk band structure of  WAs$_{2}$ with SOC.
\textbf{e}. Measured Fermi surface map of the WAs$_{2}$ at the photon energy of 35 eV.  \textbf{f}. Energy-momentum dispersion of  the [001] plane of WAs$_{2}$ along the X$_{1}$-Y-X$_{1}$ direction. \textbf{g}.  Energy-momentum dispersion of [001] plane of  WAs$_{2}$  along the X-$\Gamma$-X direction. All the data were collected at the ALS beamline 7.0.3 at a temperature of 20 K.}
\end{figure*}
\noindent
\textbf{Observation of a Dirac state} \\
\noindent
In order to illustrate the Dirac state on the [001] plane of MoAs$_{2}$, we present  detailed electronic structures of MoAs$_{2}$ [001] surface in Fig. 2.  The experimental Fermi surface as shown in Fig. 2a consists of multiple Fermi pockets indicating the metallic nature of MoAs$_{2}$.  A bridge-like line pair at the Y point  is related to the surface band at Y, which characterizes the Dirac state.  The  experimental energy contour at 360 meV binding energy in Fig. 2b constitutes the node of the Dirac cone.  A point like Dirac-node is seen in Fig. 2b, enclosed in the dotted-rectangle.  The calculated dispersion map along the  X$_{1}$-Y-X$_{1}$ direction displays the Dirac cone as shown in Fig. 2c.  First-principles calculations show the presence of a Dirac state at the Y point with the Dirac node lying at a predicted binding energy of 360 meV  below the Fermi level is in reasonable accord with the experimental results shown in Fig. 2d. Figure 2d includes photon energy dependent measurements, which provide robust evidence of a Dirac cone, despite the ARPES matrix-element effect \cite{AB1, AB2}. Photon energy dependent measurements  of  dispersion maps  along the X$_{1}$-Y-X$_{1}$ direction reveal the surface nature of the bands.  The shape of the inner linear bands forming Dirac cone does not disperse with varying photon energies, which clearly manifests the surface originated Dirac cone.  We can clearly see the surface Dirac band is surrounded by the electron-like band. It becomes stronger with the photon energy indicating a strong bulk nature. In order to illustrate the gapless cone and the  Dirac point,  the second derivative plots are presented in SI [see supplementary Fig. S1]. Therefore, dispersion maps  along the X$_{1}$-Y-X$_{1}$ confirm the presence of gapless Dirac state on the [001] surface. 

\noindent
  \textbf{Termination dependent electronic structure}\\
  \noindent
In order to demonstrate the surface dependent electronic structure in MoAs$_{2}$, we show the Fermi surface map and cuts along the high symmetry axis  on [010] surface. Figure 3a displays the Fermi surface map of the [010] plane showing a distinctly different shape compared to the [001] plane consisting of a multitude of  different Fermi pockets. Approximately circular pockets are separated by a pair of lines. Extended Fermi map consists of multiple Brillouin zones which are connected to each other forming dumb-bell shape. The rod like feature along the $\Gamma$-X direction corresponds to electron-like bands. In Fig. 3b, a calculated Fermi surface is shown which qualitatively matches with the experimental Fermi surface.  At the $\Gamma$ point, a French-curve like feature is seen in contrast to the rod like feature seen in the experimental Fermi surface  shown in Fig. 3a. Our  experimental data does not resolve all the features seen in the theoretical plot due presumably to the effects of the ARPES matrix element \cite{AB1, AB2}. Experimental  dispersion maps are taken  along the high symmetry directions in order to investigate the possible Dirac state in this plane. The maps along the H-$\Gamma$-H direction and the H$_{1}$-X-H$_{1}$ direction  have electronic structures as shown in Fig. 3c and Fig. 3d, respectively. In Fig. 3c,  a pair of electron pockets can be seen in the vicinity of the $\Gamma$ point. A hole pocket appears at a higher binding energy and does not touch the electron pocket, an obvious gap can be seen. Furthermore,  photon energy dependent dispersion maps in the Figs. 3c-d  indicate  the bulk originated bands as they disperse with photon energy. Interestingly, the gapless Dirac states are not seen in Figs. 3c-d [see also  supplementary Figs. S2-S3]. A quantum phase transition from a trivial to a non-trivial band structure can thus be seen as one moves from the [010] to the [001] surface in MoAs$_{2}$.\\
  
  \noindent
 \textbf{Electronic structure of WAs$_{2}$}\\
 \noindent 
 We investigate the effect of tuning the spin orbit coupling  in the transition metal diarsenide group by substituting Mo with W. Figure 4 shows  the transport behavior and the  detailed electronic band structure of  WAs$_{2}$ on the  [001] surface. The resistivity and magnetoresistance curves exhibit  characteristics similar to  MoAs$_{2}$. The plateau in resistivity extends nearly upto 30 K and the resistivity increases monotonically with temperature as shown in Fig. 4a. Figure 4b presents  magnetoresistance as a function of magnetic field at different temepratures. At 2 K, the magnetoresistance reaches nearly $\sim$1200\%, which is almost ten times greater than  that of the magnetoresistance of MoAs$_2$. In order to study the nature of the bulk bands in WAs$_2$, we present bulk band calculations in Figs. 4c-d.  Figure 4c shows the bulk band calculations without considering the spin-orbit coupling. It does not indicate the presence of gapless Dirac cone.  
  The inclusion of the spin-orbit  coupling, as shown in Fig. 4d, widens the gap between the linearly dispersing bands, which is consistent  with a previous study \cite{Chen}. The electronic structure of WAs$_{2}$  is studied by taking a Fermi surface map and dispersion maps along the high symmetry directions.  The Fermi surface contains multiple Fermi pockets suggesting metallic nature of the material. In order to examine the  Dirac state in WAs$_{2}$, the dispersion maps are taken along the high symmetry directions as shown in Fig. 4f-g.  Figure 4f shows the dispersion map along the X$_{1}$-Y- X$_{1}$ direction consisting of hole like bands. In contrast to MoAs$_2$, it does not possess a surface Dirac cone in  this direction.  The sharp change in the shape of the bands with the photon energy supports the bulk nature of the bands (see supplementary Fig. 4S). Similarly, the dispersion map along the X-$\Gamma$-X  direction shows  the trivial  bands and  band structures are dominated by the bulk bands as they disperse with changes in photon energy. Dirac like dispersion is absent in the dispersion maps taken along the high symmetry directions [see Fig. 4f-g].   \\

\noindent
  \textbf{Discussion }\\
  \noindent
  Our systematic study reveals the termination dependent electronic structure and effect of spin-orbit coupling in the transition metal diarsenide compounds TAs$_{2}$ (T = Mo, W). For MoAs$_2$, we study the electronic structure along the two different cleaved surfaces [001] and [010]. Both surfaces show the presence of multiple metallic Fermi pockets. Importantly, the presence of a surface Dirac cone at the Y point of the BZ is observed in the [001] plane while the [010] plane lacks such Dirac cone at any high symmetry point of the BZ. The observed Dirac point is located at about 360 meV below the Fermi level. The termination dependent Dirac point switching indicates the importance of having accurate knowledge of the projected plane  to identify the topological invariants \cite{Chris}. Furthermore, in order to study the effect of SOC, we replace Mo by a heavier element W. Interestingly, no surface originated Dirac state is observed in WAs$_2$. Therefore, our study identifies a potential surface Dirac point to trivial state transition in TAs$_2$ system. Such termination and SOC effect dependent surface Dirac switching is a unique feature observed in TAs$_2$ system. Our bulk band calculations confirm the absence of Dirac states at any momentum-energy positions in these materials, therefore, along with the fact that TAs$_2$ respects both time reversal and inversion symmetry, it negates the possibility that the TAs$_2$ is either an intrinsic Dirac semimetal or  a Weyl semimetal. \\
  In conclusion, we  perform ARPES measurement in conjunction with first principles calculations in TAs$_{2}$ (T = Mo, W). Our study provides an evidence of switching of the Dirac state with the cleavage plane in MoAs$_{2}$. On the other hand, WAs$_{2}$ does not show the indication of the Dirac state. Our report highlights the importance of cleavage plane and spin-orbit coupling to understand the Dirac states in transition metal diaresenides. Moreover, this study provides a new platform to investigate the intriguing phenomena associated with the cleavage plane in transition metal dipnictides. 
   

 \noindent 
\textbf{ METHODS}\\
\noindent
\textbf{Sample growth and characterization}\\
The single crystals of  MoAs$_{2}$ and WAs$_{2}$ were grown using the chemical vapor transportation \cite{J}. The crystals form needlelike structures with dimensions $\sim$ 1$\times$0.8$\times$0.4 mm$^{3}$. The chemical composition was
proven by energy-dispersive X-ray analysis using a FEI scanning electron microscope equipped with an EDAX
Genesis XM4 spectrometer. The crystal structure was examined at room temperature on a Kuma-Diffraction KM4
four-circle X-ray diffractometer equipped with a CCD camera using Mo K$\alpha$ radiation. The experiment confirmed the monoclinic crystal structure provided the lattice constants concurring with the previously reported values \cite{J}.  Electrical resistivity measurements were performed
in the temperature range of 2 - 300 K employing a conventional four-point ac technique implemented in a Quantum
Design PPMS platform. The electrical contacts were made using silver epoxy paste. The  perpendicular magnetic field was applied to measure the magnetoresistance.\\

\noindent
\textbf{ARPES measurements}\\
\noindent
 Synchrotron-based ARPES measurements of the electronic structure were performed at the Advanced Light Source (ALS), Berkeley at beamlines 4.0.3, 7.0.2 and  10.0.1 equipped with high-efficiency R8000, R4000, and R4000 electron analyzers, respectively. The energy resolution was set to be better than 20 meV for the measurements with the synchrotron beamline. The angular resolution was set to be better than 0.2$^{\circ}$ for all synchrotron measurements. Samples were cleaved in situ and measured at 15 - 25 K in a vacuum better than 10$^{-10}$ torr. They were found to be very stable and without degradation for the typical measurement period of 20 hours.\\
 
 \noindent
\textbf{Calculations}\\
\noindent
 The electronic structure was calculated by using Vienna Ab Initio Simulation Package (VASP) \cite{VASP1} based on the density functional theory (DFT) \cite{DFT} with the projector augmented wave (PAW) method \cite{VASP1,VASP2}. We used the generalized-gradient approximation (GGA) \cite{GGA} to model the exchange-correlation effects. The energy cutoff of the plane wave basis used was 400 eV, and the size of  k-mesh was $ 16 \times 16 \times 7$. The tight-binding model Hamiltonian was obtained by projecting first-principles results onto Wannier orbitals using the VASP2WANNIER90 \cite{WANNIER90}. The Mo \textit{d} and As \textit{p} orbitals were included to get the band structure around the Fermi level. We calculated the surface states by using iterative Green's function method with the WannierTools package \cite{Wanniertool}.\\

\noindent
\textbf{DATA AVAILABILITY}\\
\noindent
The data that supports the findings of this study are available from the corresponding author upon reasonable request.\\ 

\noindent
\textbf{ACKNOWLEDGEMENT}\\
M.N. is supported by the Air Force Office of Scientific
Research under award number FA9550-17-1-0415 and
the National Science Foundation (NSF) CAREER award
DMR-1847962.  The work at Northeastern
University was supported by the US Department
of Energy (DOE), Office of Science, Basic Energy
Sciences grant number DE-FG02-07ER46352, and
benefited from Northeastern University’s Advanced
Scientific Computation Center (ASCC) and the NERSC
supercomputing center through DOE grant number
DE-AC02-05CH11231. We thank Sung-Kwan Mo, Jonathan Denlinger, Eli Rotenberg, Aaron Bostwick and Chris Jozwiak for beamline assistance at the LBNL.\\

\noindent 
\textbf{AUTHOR CONTRIBUTIONS}\\
M.N. and D. K. conceived the study; D.K. synthesized the samples and performed the electrical transport characterization; G.D. performed the measurements with the help of M.M.H., K.D.,  F.K., C.S., S.R.,  and M.N.; G.D. and M.N. performed the data analysis and figure planning; W.-C.C., B.S., B.W., and A.B. performed the ab initio calculations and topological analysis; A.B. was responsible for the theoretical research direction; G.D. and M.N. wrote the manuscript with input from all authors. M.N. was responsible for the overall research direction, planning, and integration among different research units.\\

\noindent
Correspondence and requests for materials should be addressed to M.N. (Email: Madhab.Neupane@ucf.edu).
 \bigskip


\end{document}